\documentclass{PoS}

\title{Status and potentialities of the JUNO experiment}

\ShortTitle{Status and potentialities of the JUNO experiment}

\author{{V. Antonelli}\thanks{On behalf of the JUNO Collaboration.}\\
        I.N.F.N., Sezione di Milano and Dipartimento di Fisica, Universit\'a degli Studi di Milano,\\
        Via Celoria 16, 20133-Milano (IT)\\
        E-mail: \email{vito.antonelli@mi.infn.it}}

\author{L. Miramonti*\\
        I.N.F.N., Sezione di Milano and Dipartimento di Fisica, Universit\'a degli Studi di Milano,\\
        Via Celoria 16, 20133-Milano (IT)\\
       E-mail: \email{lino.miramonti@mi.infn.it}}

\abstract{One of the main open issues of neutrino physics is the determination of the mass hierarchy, discriminating between the two possible 
ordering of the mass eigenvalues, known as Normal and Inverted Hierarchies. The solution of this puzzle would have a significant impact both on elementary particle 
physics and astrophysics. A possible way to investigate the problem is the study, with medium baseline reactor antineutrinos, of the mass dependent corrections
to inverse $\beta$ decays. This is the idea pursued by JUNO, a multipurpose underground liquid scintillator experiment that will start data taking in very few years from
now. The main characteristics and the status of the experiment is discussed, together with its rich physics program. We focus in particular on the potentiality 
for mass hierarchy determination, the main goal of the experiment, oscillation parameters accurate measurements, supernova and solar neutrinos and geoneutrinos 
study.}

\FullConference{XVII International Workshop on Neutrino Telescopes\\
		13-17 March 2017\\
		Venezia, Italy}

\begin{document}

\section{Introduction: the mass hierarchy determination and the JUNO option}
\label{introduction}
\begin{figure}[h]
\includegraphics[scale=0.20,angle=270]{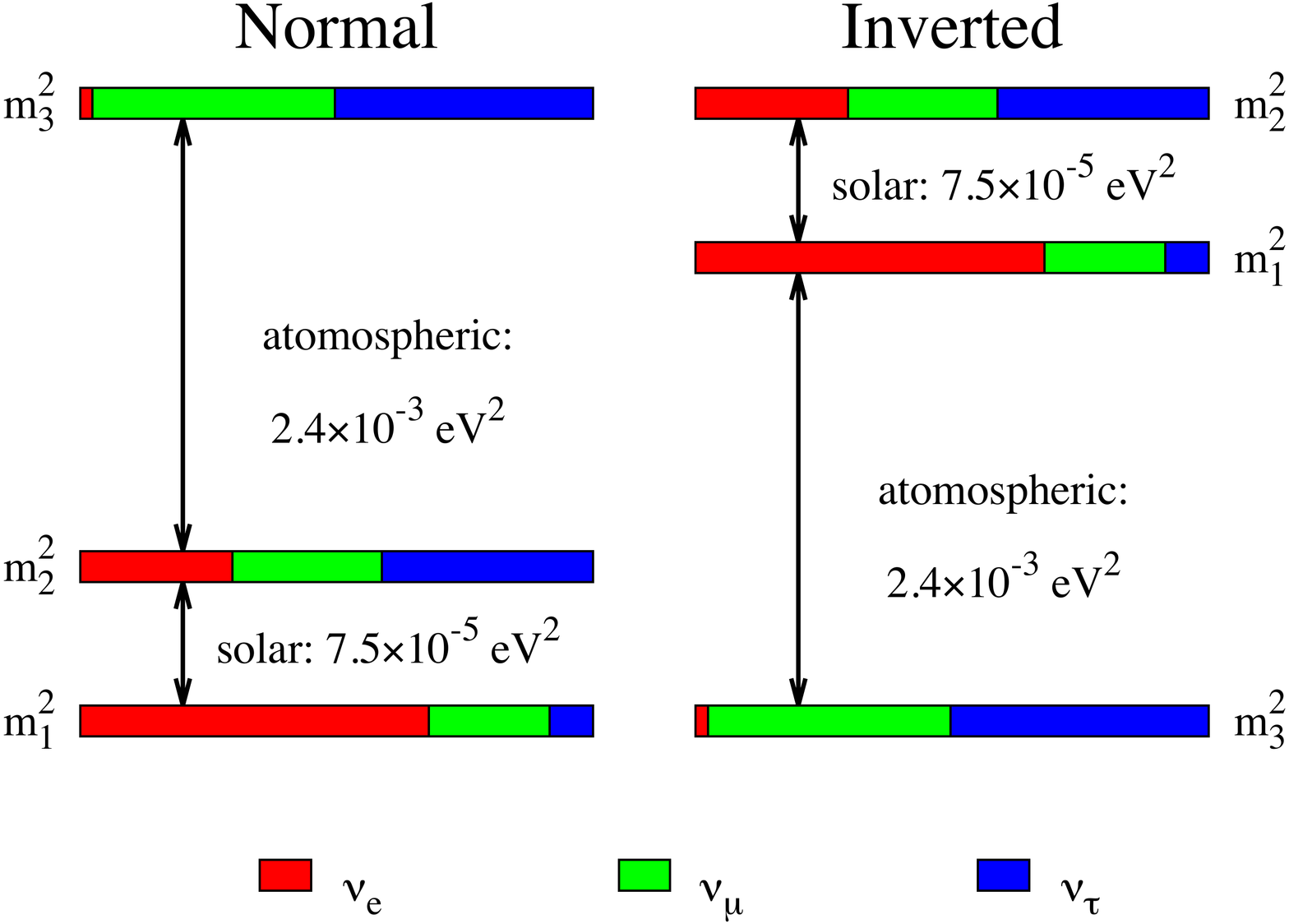}
\caption{Neutrino mass eigenstate flavor composition and mass pattern in the two cases of normal (left) and inverted (right) hierarchies. Taken 
from~\cite{JUNO-Yellow-Book}.}
\label{hierarchy}
\end{figure}

Almost one century after its introduction by Pauli, 
neutrino is still one of the most mysterious and intriguing elementary particles. It took almost half a century to prove that it is massive and oscillating, but we don't know yet its real nature (Dirac or Majorana fermion)
and we don't have neither a clear idea of the value of its mass, nor a unique explanation of the reason for which it is so much lighter than all the other particles. The answers to these long standing questions would have a great impact both on elementary particle physics and on astrophysics.
A first step forward 
could be the discovery of the neutrino mass eigenvalues hierarchy. 
The two possible scenarios still compatible with the 
data by different classes of neutrino experiments, usually denoted as normal and inverse hierarchy (NH and IH), are represented in fig.\ref{hierarchy}. 
In the first case the third neutrino mass eigenvalue would be the highest one and $|\Delta m^2_{31}| = |\Delta m^2_{32}| + \Delta m^2_{21}$, 
where we denoted by $\Delta m^2_{ij} = m_i^2 - m_j^2$ the differences of the squared mass eigenvalues.
In case of inverted hierarchy, instead, $m_3$ would be the lowest eigenvalue and the relation would be
$|\Delta m^2_{31}| = |\Delta m^2_{32}| - \Delta m^2_{21}$. 

Hierarchy determination is essential both for model building (discriminating between different possible extensions of the Standard Model) and for the evaluation of the discovery potential of present and future experiments, like the ones looking for leptonic CP violation or for neutrinoless double $\beta$ decays ($0\nu2\beta$). 
The decay probability for $0\nu2\beta$ is proportional to 
the ``effective mass'' $<m_{\beta\beta}>$, a combination of neutrino mass eigenvalues weighted by the squares of the mixing matrix elements (mixing angles $\theta_{ij}$ and Majorana phases 
$\phi_i$), that, in the 3 flavor scenario, is given by:
\begin{eqnarray}
< m_{\beta \beta} > & = & \left|\sum_{i=1}^N U_{ei}^2 \, m_i\right|
 = \left|cos^2 (\theta_{13}) (m_1 \, e^{i \phi_1} \, cos^2 \theta_{12} + m_2 \, e^{i \phi_2} \, \sin^2 \theta_{12}) + m_3 \, sin^2 \theta_{13}\right| 
\end{eqnarray}
For IH the possible $\langle m_{\beta \beta} \rangle$ value  could reach the level of a few tenths of meV, which should be accessible by the next-generation experiments; on the opposite, in case of NH the effective neutrino mass would be at least one order of magnitude lower and much more difficult to test experimentally~\cite{doublebeta}. 
The situation changes significantly in case of more complicated models including sterile neutrinos (like 3+1 models), as discussed, for instance, in~\cite{double-beta-sterile}. 
Recent data from long baseline accelerator experiments~\cite{LBL-hierarchy} and from  cosmology \cite{cosmology-hierarchy} seem to favor NH. 
However, the theoretical  interpretation and the statistical significance of  the conclusions on the mass hierarchy derived from these results are not univocally 
determined~\cite{discussion-hierarchy} and they suffer from uncertainty related to other assumptions or parameters (adopted cosmological model, value of CP violation, etc.). 

\begin{figure}[h]
\centering
\includegraphics[width=70mm]{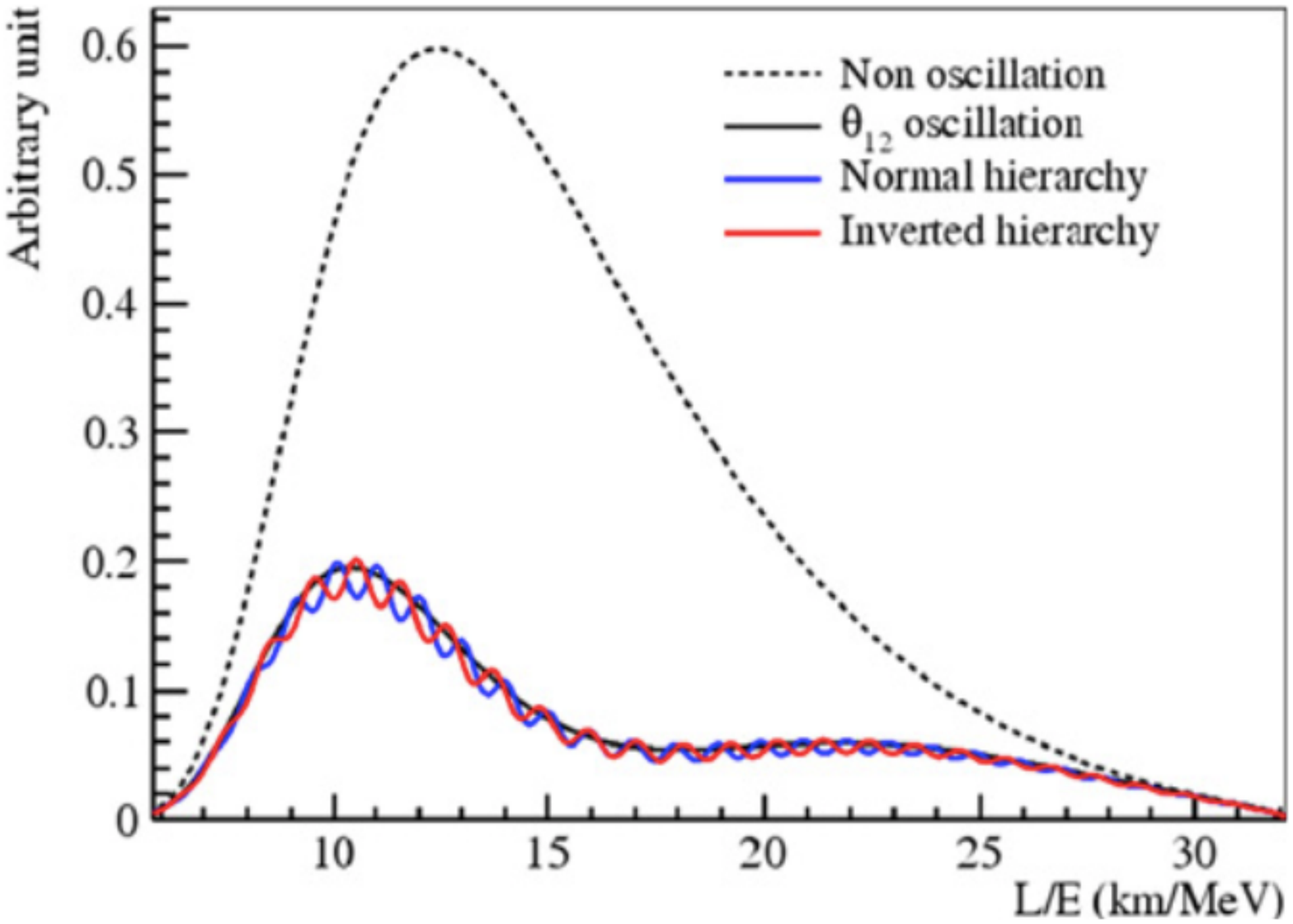}
\qquad\qquad
\includegraphics[height=45mm]{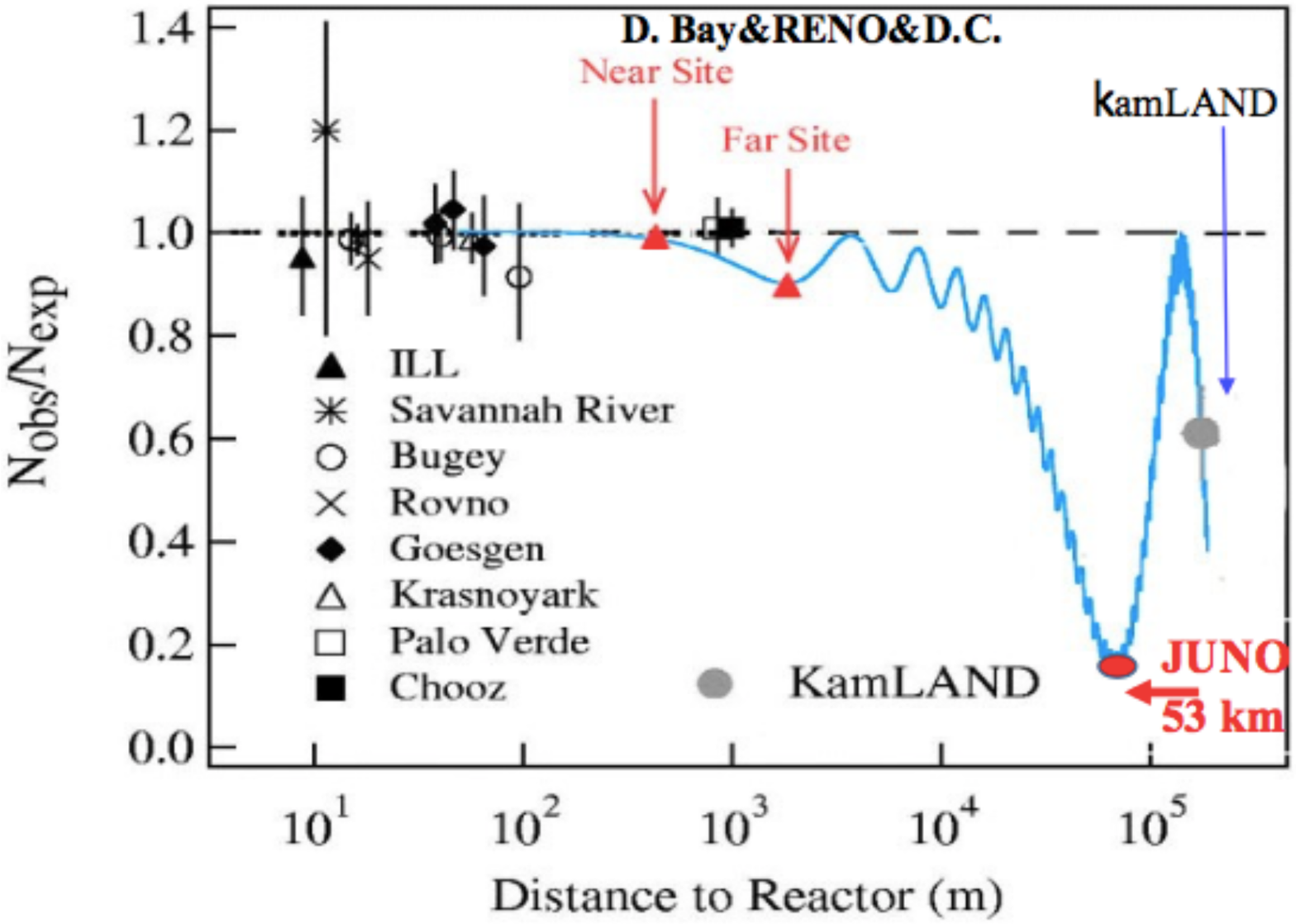}
\caption{Left panel: Reactor $\bar{\nu}$ IBD spectrum as a function of L/E, in absence of oscillation and in the oscillation case for NH and IH. 
Taken from~\cite{JUNO-Yellow-Book}, where it is reprinted with permission from~\cite{paper-medium-baseline}. 
Right: Ratio between the IBD events and the expected number in case of no oscillation, as a function of the baseline.}.
 \label{fig-spectrum-baseline}
\end{figure}

The significantly different from zero value of  $\theta_{13}$, the mixing angle between the electron flavor and the third generation of neutrino mass eigenstates, 
 makes possible the realization of the idea~\cite{Petcov-et-al} to extract 
 the mass hierarchy in a completely different way, from the study, with medium baseline reactor antineutrinos, of the hierarchy dependent oscillation probability corrections 
 proportional to $sin^2 \theta_{13}$. The ($\bar{\nu}_e$) survival probability can be written as: 
\begin{eqnarray}
P_{ee} & =  &1 - cos^4 (\theta_{13}) \, sin^2 (2 \theta_{12}) \, sin^2 \frac{\Delta m^2_{21} L}{4 E} - sin^2 2 \theta_{13} \, \left(cos^2 (\theta_{12})\, 
sin^2 \frac{\Delta m^2_{31} L}{4 E} + sin^2 (\theta_{12})  \, sin^2 \frac{\Delta m^2_{32} L}{4 E}\right) 
\nonumber\\ 
&= &1 - cos^4 (\theta_{13}) \, sin^2 (2 \theta_{12}) \, sin^2 \frac{\Delta m^2_{21} L}{4 E} - P_{MH}
\label{Pee}
\end{eqnarray}
In the last expression we indicated as $P_{MH}$ the mass hierarchy dependent contribution to the 
oscillation probability (proportional to $sin^2 (2 \theta_{13})$, that can be rewritten in the following form:
\begin{equation}
P_{MH} = \frac{1}{2} \, sin^2 2 \theta_{13}\, 
\left(1 -\sqrt{1 - sin^2 (2 \theta_{12}) \sin^2(\frac{\Delta m^2_{21} L}{4 E})}\,  \, 
cos \left(2 \left|\frac{\Delta m^2_{ee} L}{4 E}\right| \pm \phi \right) \right)\, , 
\label{PMH}
\end{equation}
where $\Delta m^2_{ee}$ represents the combination 
$ \Delta m^2_{ee} = \left(cos^2 (\theta_{12}) \Delta m^2_{31} + sin^2 (\theta_{12}) \Delta m^2_{32}\right)$ and we denoted by 
$\phi$ a new quantity, defined in such a way that $sin \, \phi$ and $cos \, \phi$ are combinations of the mass and mixing parameters of 
the 1-2 sector\footnote{For the details of the calculation see~\cite{JUNO-Yellow-Book}}.
The sign in front of $\phi$ is +1 for NH and -1 for IH. Hence the survival (or oscillation) probability and, consequently, the observed spectrum are characterized by the presence of fastly oscillating terms, in opposition of phases for the two cases of NH and IH, superimposed to the general oscillation pattern valid for both hierarchies. 
A spectrum study with a good energy resolution is needed to discriminate between the two possible hierarchies.  
The golden channel for a reactor antineutrino beam is the inverse $\beta$ decay: 
$\bar{\nu}_e + p \to e^+ + n$. 
The expected spectrum presents the structure described 
above and represented in the left panel of fig.~\ref{fig-spectrum-baseline}, where one can note that the hierarchy dependent corrections are in opposition of phases for the two cases of NH and IH.
To study mass hierarchy  is convenient to maximize the oscillation effect and fig.\ref{fig-spectrum-baseline}  (right) 
shows that this is obtained for about 50 km of baseline. This is the case of JUNO experiment. 

\section{The JUNO detector and experimental site and the status of the experiment}

JUNO (Jiangmen Underground Neutrino Observatory) is a multipurpose underground reactor antineutrino experiment, under construction near Kaiping, 
in the South of China~\cite{JUNO-Yellow-Book,JUNO-technical}. JUNO collaboration includes 71 institution members, plus 2 observers, spread over 3 continents. 
The reactor $\bar{\nu}_e$ flux under investigation at JUNO will arrive mainly by 10 (in the original project) nuclear cores\footnote{There is the possibility that two nuclear cores will not be available yet when JUNO will start data taking} settled in the two different power plants 
Yiangijiang (nominal power of 17.4 GW) and Taishan (18.4 GW), with a baseline between 52 and 53 km; a non negligible 
contribution will come also from the Daya Bay and the Huizhou cores located at a distance of about 200 km. 
The JUNO detector is located underground, with a rock overburden of about 700 m ($\sim 1900 {\rm m.w.e.}$), in order to reduce the cosmic background. 
It is made up by a very thick (120 mm) acrylic sphere of 35.4 m of diameter, containing 20 kton fiducial mass of liquid scintillator, 
Linear Alkylbenzene  (LAB), chosen for its very good properties of optical transparency, high flash point, low chemical reactivity and, mainly, excellent light yield. The liquid scintillator also contains 3 g $l^{-1}$ of 2,5-Dyphenyloxazole (PPO) and 15 mg $l^{-1}$ of 
p-bis(O-methylstyril)-benzene (bis-MB), a wavelength shifter. 
The acrylic sphere is sustained by a steel structure and surrounded by a double system of photomultipliers 
and it is submerged in an external water tank, acting as muon \v{C}erenkov veto, to protect from the natural radioactivity by
the surrounding rock and the air.
To enable an accurate measurement of muons directions, a muon tracker will be installed on the top of the water pool, using the plastic 
scintillator strips decommissioned from the target tracker of the OPERA experiment. 

The main reaction under investigation at JUNO is the inverse $\beta$ decay (IBD) of antineutrino on proton. This kind of events has a clear experimental signature, characterized by the coincidence between the prompt emission of a couple of photons of 511 keV, coming from the annihilation of the positron produced in the IBD final state, and the delayed emission of another $\gamma$  of characteristic 
energy (2.2 MeV), due to the absorption of the second final state particle, the neutron. 
The IBD cross section has an energy threshold $E_{\nu} \geq 1.8 \, {\rm MeV}$ and the spectrum is contained between 1.8 and 8-9 MeV and 
presents a maximum for $\bar{\nu}_e$ energy around 4 MeV, corresponding, for JUNO, to $L/E \sim13$. 
As shown in the left graph of fig.~\ref{fig-spectrum-baseline}, the effect of oscillation, in addition to the reduction of the number of events, is that of 
modifying partially the spectrum shape (due to the smallness of $\Delta m^2_{21}$, the larger values of $\frac{L}{E}$ are more 
suppressed) and creating the wiggles (described in previous section) due to the fast oscillating terms dependent on the mass hierarchy. 
To analyze the spectrum, distinguishing the wiggles position and extracting the desired information on mass hierarchy, is an hard task, but JUNO can take advantage by the optimal baseline choice, by a very high statistics (made possible by the large detector mass and the proximity to several reactors) and, above all, by a very good energy resolution, which will be the real key factor for the success of the experiment.The energy dependence of the resolution is deeply investigated and it can be expressed by~\cite{JUNO-Yellow-Book}:
\begin{equation} 
\frac{\sigma_E}{E} = \sqrt{\left(\frac{a}{\sqrt{E}}\right)^2 + b^2 + \left(\frac{c}{E}\right)^2} 
\simeq 
\sqrt{\left( 
\frac{a}{\sqrt{E}}
\right)^2
+ 
\left(
\frac{1.6 \, \, b}{\sqrt{E}}
\right)^2 +
\left(
\frac{c}{1.6 \, \sqrt{E}}
\right)^2 
}
\label{E-resolution}
\end{equation} 
In the two formulas[~\ref{E-resolution}] the first term, proportional to the coefficient $a$, express the stochastic contribution and the rest is due to non 
stochastic contribution. 
The requirement that should be respected by the experiment is a total energy resolution $\frac{\sigma_E}{\sqrt{E}} \leq 3 \%$ and, 
hence, $\sqrt{a^2 + \left(1.6 \, \, b \right)^2 + (\frac{c}{1.6})^2} \leq 3 \%$\,. 
The very good energy resolution and high photon yield, together with the already cited huge mass and the extremely high coverage (around $80 \%$), are among the main technical characteristic representing the strong points of JUNO~\cite{JUNO-technical}, as 
shown in table~\ref{JUNO-performances}, comparing JUNO performances with the ones of other important liquid scintillator neutrino experiments. 
\begin{table}[h]
\begin{tabular}{l|c|c|c|r}
\hline
Experiment  & Day Bay & Borexino & KamLAND & JUNO\\  \hline
Liquid Scintillator mass  & 20 ton & $\sim$ 300 ton & $\sim $1 kton  & 20 kton\\ \hline
Coverage &  $\sim 12\%$ & $\sim 34\%$ & $\sim 34\%$ & $\sim 80\% $\\ \hline
Energy Resolution & $\frac{7.5 \%}{\sqrt{E}}$ & $\frac{\sim 5 \%}{\sqrt{E}}$ & $\frac{\sim 6 \%}{\sqrt{E}}$ & $\frac{\sim 3 \% }{\sqrt{E}}$\\ \hline
Light Yield & $\sim 160 \, {\rm \frac{p.e.}{MeV}}$ &  $\sim 500 \, {\rm \frac{p.e.}{MeV}}$ & $\sim 250 \, {\rm \frac{p.e.}{MeV}}$ & 
$\sim 1200 \, {\rm \frac{p.e.}{MeV}}$ \\
\hline
\end{tabular}
\caption{Summary of the JUNO values for some of the main techincal parameters, compared with other famous neutrino experiments using 
liquid scintillators. Taken from~\cite{JUNO-Yellow-Book} and~\cite{JUNO-technical}\, .
}
\label{JUNO-performances}
\end{table}
In order to reach this ambitious goals at JUNO, a particular attention has been paid to the photomultiplier system.  The detector will be 
surrounded by 17000 large (20") photomultipliers (PMTs), among which 25000 other smaller (3") PMTs will be interposed. 
This double calorimetry system guarantees a powerful event reconstruction and it offers the possibility of an internal cross check that should 
help in reducing the non stochastic uncertainty and improving the high precision oscillation parameter measurement. In addition to this, the 
possibility of a double readout will be useful for unbiased energy and rate measurements, important, for instance, for the supernova neutrino studies.
Two kind of large PMTs will be used, realized by two different producers: the Chinese Company North Night Vision Technology (NNVT) and Hamamatsu. 
The tests performed gave very encouraging results for both types of large PMTS. For a detailed discussion on 
PMTs requirements, innovative design, development and performances, we refer the interested reader to~\cite{talk-Gioacchino} and to the contribution to this Conference 
of Yifang Wang~\cite{talk-Yfang}. Here we just recall the very good values of the Quantum Efficiency (around $30\%$ for both kinds of PMTs) and of the Relative Detection Efficiency ($100\%$ for Hamamatsu and even better, $110\%$, for the Chinese PMTs), essential to reach the desired 
energy resolution. Also the Transient Time Spread, important to identify the vertex, is satisfactory (12 ns for the NNVT and even lower, 3 ns, for the Hamamatsu PMTs).
 
Important goals have been reached also for another delicate aspect of the experiment, the optical transparency and the radiopurity of the liquid scintillator. A pilot plant 
has been installed at Daya Bay, to test the overall design and the efficiency of different purification systems: distillation, steam stripping, water extraction and 
use of $Al_2O_3$\,. The results of the test have been positive.
The technological difficulties of realization of the huge central detector structure (around 600 tons of stainless steel truss and an analogous mass of very 
thick acrylic panels), in  particular for the shrinkage and shape variations of the acrylic panels, have been faced and solved and the radioactivity level of the panels have been checked and it is under control.  

\section{Milestones of the analysis}
The primary goal of JUNO experiment is that of investigating the mass hierarchy and, as already explained, this will be done by analyzing the spectrum 
of the antineutrino inverse $\beta$ decay on proton. The experimental spectrum will be compared with the theoretical one, which is a function of the oscillation parameters (neutrino squared mass differences and mixing angles) and of the mass hierarchy. These data will enter in a global $\chi^2$ fit,
including also the outputs of the other neutrino experiments, constraining the oscillation parameters.
The discrimination of the two possible mass hierarchies is based on the comparison between the values of the $\chi^2$ corresponding to the best fit 
 points of the two different hierarchies and the difference between these two $\chi^2$ values, 
 $\Delta \chi^2_{MH} = \left|\chi^2_{MIN} (NH) - \chi^2_{MIN} (IH) \right|$ is a measure of the ``mass hierarchy sensitivity'', that is the discrimination 
 power of the experiment~\cite{JUNO-Yellow-Book}. The technical aspects of the analysis that will be possible to perform with the real experimental 
 data, the statistical interpretation of the results and the related subtelties are subject of wide discussion in literature~\cite{theory-chi2}. An alternative way 
 to recover the mass hierarchy from medium baseline reactor experiments is also discussed in~\cite{Petcov-Bilenky}.
 
\begin{figure}
 \includegraphics[scale=0.20]{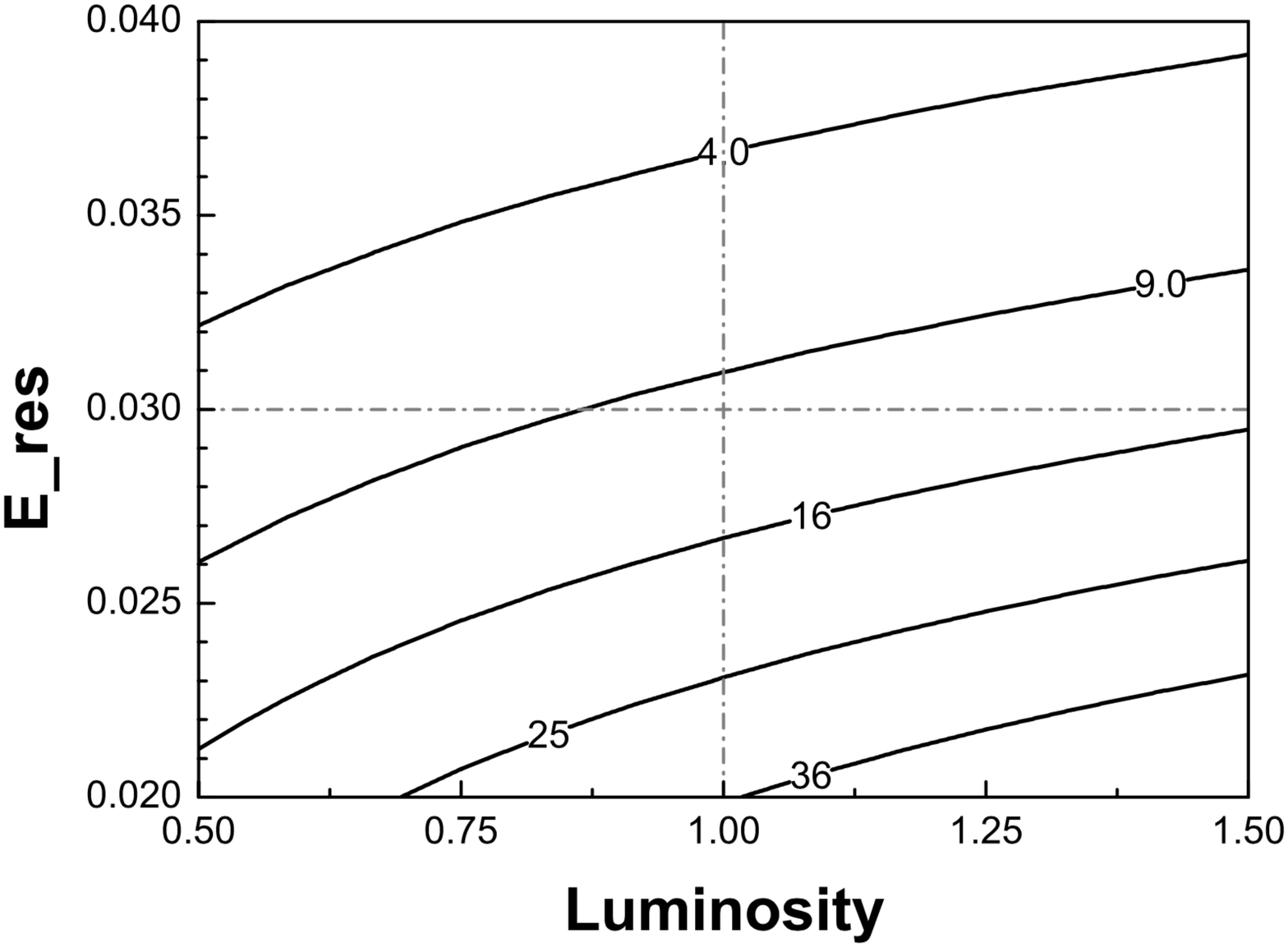}
 \caption{Iso-$\Delta \chi^2_{MH}$ contour lines as a function of the number of events (indicated as ``luminosity'') and of the energy resolution. 
 Luminosity 1 corresponds to the expected statistics after 6 years of data taking with JUNO nominal luminosity and a detector efficiency of 
 $80 \%$.}
 \label{fig-deltachi2}
 \end{figure}

 The crucial point for the success of the analysis (in addition to the high sensitivity) is, clearly, the energy resolution. In fig.~\ref{fig-deltachi2}, the 
 hierarchy sensitivity is represented as a function of the number of data taken and of the energy resolution. It is noteworthy that, for a resolution equal 
 $\frac{3\%}{\sqrt{E}}$, after six years of data taking at the nominal luminosity of the experiment and with an estimated detection efficiency of $73\%$,
 it should be possible to reach a value of $\Delta \chi^2_{MH}$ around 10. 
 
  In the derivation of these results, we already took into account  some secondary effects which enter in the analysis and reduce the hierarchy sensitivity with respect 
  to the ideal case. 
 The main one is the impact of the differences in the baselines for the different nuclear cores from which the reactor antineutrinos originate\,.
 The net effect is estimated to be a degradation of about 4-5 
 units of the $\Delta \chi^2_{MH}$ value, to which one should add a further reduction of 1-2 units due to the reactor shape and statistical background uncertainties. 
On the other hand, this is essentially compensated  by another effect, acting in the opposite direction, of increasing the 
$\Delta \chi^2_{MH}$ value.   
 By the time at which the JUNO data analysis will performed, we should have at our disposal additional information (coming mainly by long baseline accelerator neutrino experiments) on an important parameter influencing the global statistical analysis, that is $\Delta m^2_{\mu\mu}$. This is the combination (different  from the one entering in formula \ref{PMH}) of the different $\Delta m^2_{ij}$ which is relevant for the experiments studying the muonic neutrino disappearance. It has been estimated\ that a measurement ``external'' to JUNO of $|\Delta m^2_{\mu\mu}|$ with $1\%$
 accuracy would bring to an increase of  $\Delta \chi^2_{MH}$ ranging from 4 to 12 units.\\
 Hence, considering all the aspects of the problem, it is realistic to expect that, after six years of data taking, JUNO could discriminate the mass hierarchy 
 at about $4 \sigma$ of confidence level. JUNO result, differently from the LBL accelerator one,  will not suffer from any ambiguity on the CP violation phase and 
 from the uncertainty on the Earth density profile (because JUNO essentially studies the oscillations in vacuum). 
Moreover, it will depend only mildly on the eventual presence of sterile neutrinos.

\section{The JUNO Physics program }
JUNO is a multipurpose experiment and, in addition to the mass hierarchy determination, it will perform other measurements and analyses, making its physics program very rich. We comment the most significant ones, referring the interested reader to~\cite{JUNO-Yellow-Book} for a more exhaustive discussion. 
\subsection{The mass and mixing parameters determination}
Thanks to the very high statistics and the excellent energy resolution JUNO will make possible a significant improvement in the precision measurement, at the subpercent level, of three mass and mixing parameters, as summarized in table~\ref{table-parameters}. For the present values of the oscillation parameters accuracy we based our analysis mainly on the two global fits of~\cite{Capozzi-Lisi-plus-Esteban-et-al}.

\begin{table}[h]
\begin{centering}
\begin{tabular}{c|c|c|c}   
Oscillation parameter & Current accuracy & Dominant experiment(s) & JUNO potentialities \\
 & (Global $1\sigma$) & &  \\ \hline
 $\Delta m^2_{21}$ &  $2.3 \%$ & KamLAND & $0.59 \%$ \\ \hline
 $\Delta m^2 = \left| m_3^2 - \frac{1}{2} (m_1^2 + m_2^2 ) \right|$ & $1.6 \%$ & MINOS, T2K & $0.44 \%$ \\ \hline
 $sin^2 \theta_{12}$ & $\sim 4-6 \%$ & SNO & $ 0.67\%$\\ \hline
\end{tabular}
\end{centering}
\caption{Comparison of the present accuracy and the JUNO potentialities for 3 oscillation parameters.}
\label{table-parameters}
\end{table}   

\subsection{Supernova neutrinos}

A detector like JUNO can perform interesting studies of supernovas, both with the ``direct'' measurement  
of the neutrino burst produced by the possible collapse of a next nearby supernova (SN) and with the 
study of the diffuse supernova background.

\begin{figure}
\vspace{-3.5 truecm}
\hspace{-0.25 truecm}
\includegraphics[scale=0.52,angle=270]{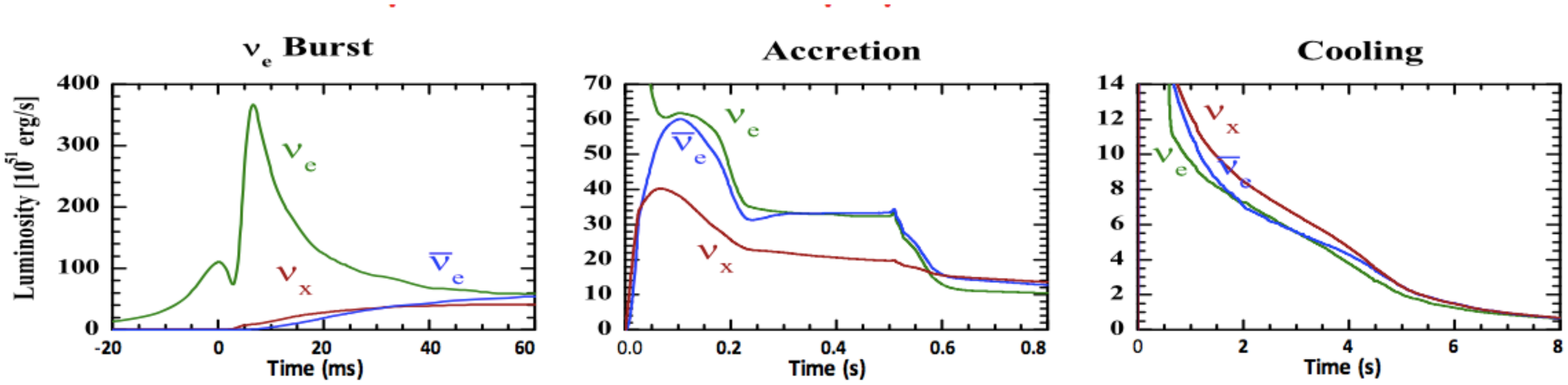}
\vspace{-4truecm}
\caption{Representation of the different flavor neutrinos and antineutrinos emission in the 3 different phases of a core-collapse SN emission.
Taken from~\cite{JUNO-Yellow-Book}}
\label{fig-SN}
\end{figure}

The explosion of a Supernova (SN) causes the emission in neutrinos of a huge amount of energy over a long time range. 
Many parameters of interest for elementary particle physics and for astrophysics can be studied by means of the analysis of 
all the three different stages in which the process can be divided: the very fast initial neutrino burst (lasting for a few tenths of 
ms), the ``accretion'' and the ``cooling'' phases. These phases offer complementary information because they are characterized 
by emission spectrums quite different among themselves, for what concerns the neutrino (antineutrino) flavor composition and the luminosity, as 
shown in fig.~\ref{fig-SN}.
The table~\ref{tab-sn-channels} reports the main interaction channels that can be investigated to study the 
neutrino emission by a supernova and the number of expected events in JUNO for a SN distance of 10 kpc,
a typical distance of astronomical relevance. One can notice the high number of events (about 5000) for the main 
channel  (the inverse $\beta$ decay) and the fact that the expected statistics is quite significant (of the 
order of hundreds of events) also for various other interaction channels.      

\begin{table}[h]
\begin{centering}
\begin{tabular}{c|c|c}
Process & Type & Events $\left( \langle E_{\nu}  \rangle= 14 {\rm MeV} \right)$ \\ \hline
$\bar{\nu}_e + p \to e ^+ + n$ & CC &  $5.0 \times 10^3$\\\hline
$\nu + p \to \nu + p $ & NC & $1.2 \times 10^3$\\\hline
$ \nu + e^- \to \nu + e^-$ & ES & $3.6 \times 10^2$\\\hline
$\nu + ^{12}C \to \nu + ^{12}C^*$ & NC & $3.2 \times 10^2$\\ \hline
$\nu_e + ^{12}C  \to e^- + ^{12}{\rm N}$  & CC & $0.9 \times 10^2$\\ \hline
$\bar{\nu_e} + ^{12}C  \to e^+ + ^{12}{\rm B}$ & CC & $1.1 \times 10^2$\\ 
\end{tabular}
\caption{Expected number of events in JUNO, for the main channels for neutrinos produced by a SN at a distance of 10 kpc. Taken from~\cite{JUNO-Yellow-Book}.}
\end{centering}
\label{tab-sn-channels}
\end{table}

\subsection{Solar neutrinos}

\begin{figure}[h]
\includegraphics[scale=0.24]{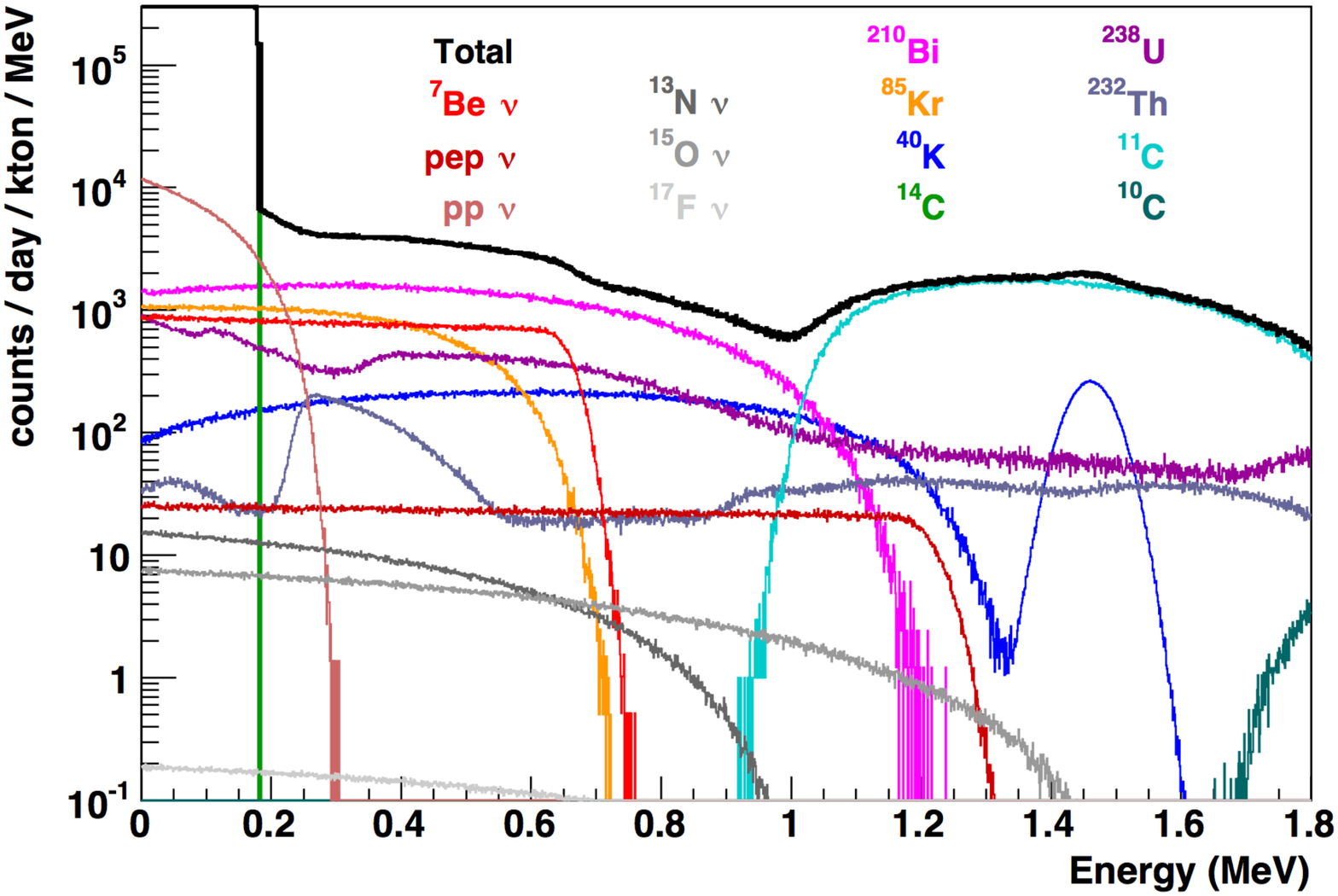}
\caption{Expected  low energy solar neutrino signal, represented  together with the main background contributions relevant in this region, for the case 
of typical required radiopurity level (for instance $1 \times 10^{-16} {\rm g \, g^{-1}}$ for $^{238} {\rm U}$ and $^{232} {\rm Th}$), approximately analogous 
to that of KamLAND solar phase purity level. Taken from~\cite{JUNO-Yellow-Book}.}
\label{fig-Be7}
\end{figure}

The neutrinos emitted by our star can be studied at JUNO, with the concrete possibility of giving a 
contribution to the solution of some of the main puzzles of this field, by taking advantage from the main strengths of the detector, that is 
the high mass of the liquid scintillator (that guarantees an high statistics) and the very good energy resolution. A key parameter 
for the success of these studies will be the radiopurity level that will be reached. This is particularly true for the analysis
of the $^7 {\rm Be}$ contribution, which require the capability to disentangle the signal from the one of different natural background sources 
relevant in the sub-MeV region. The main ones are the neutrinos emitted in the decay chains of the radioactive isotopes 
$^{238} {\rm U}$,  $^{40} {\rm K}$, $^{210} {\rm Bi}$ and $^{85} {\rm Kr}$, as shown in fig.~\ref{fig-Be7}.

An improvement in the accuracy of the determination of the flux for $^7Be$ and for $^8 B$, the other contribution to the solar neutrino 
spectrum that can be studied at JUNO, could help in the search for the discrimination between the different possible versions (low Z and
high Z) of the Standard Solar Model  and the solution of the so called ``metallicity problem''~\cite{metallicity-e-SSM}. 

\begin{figure}[h]
\centering
\includegraphics[height=40mm]{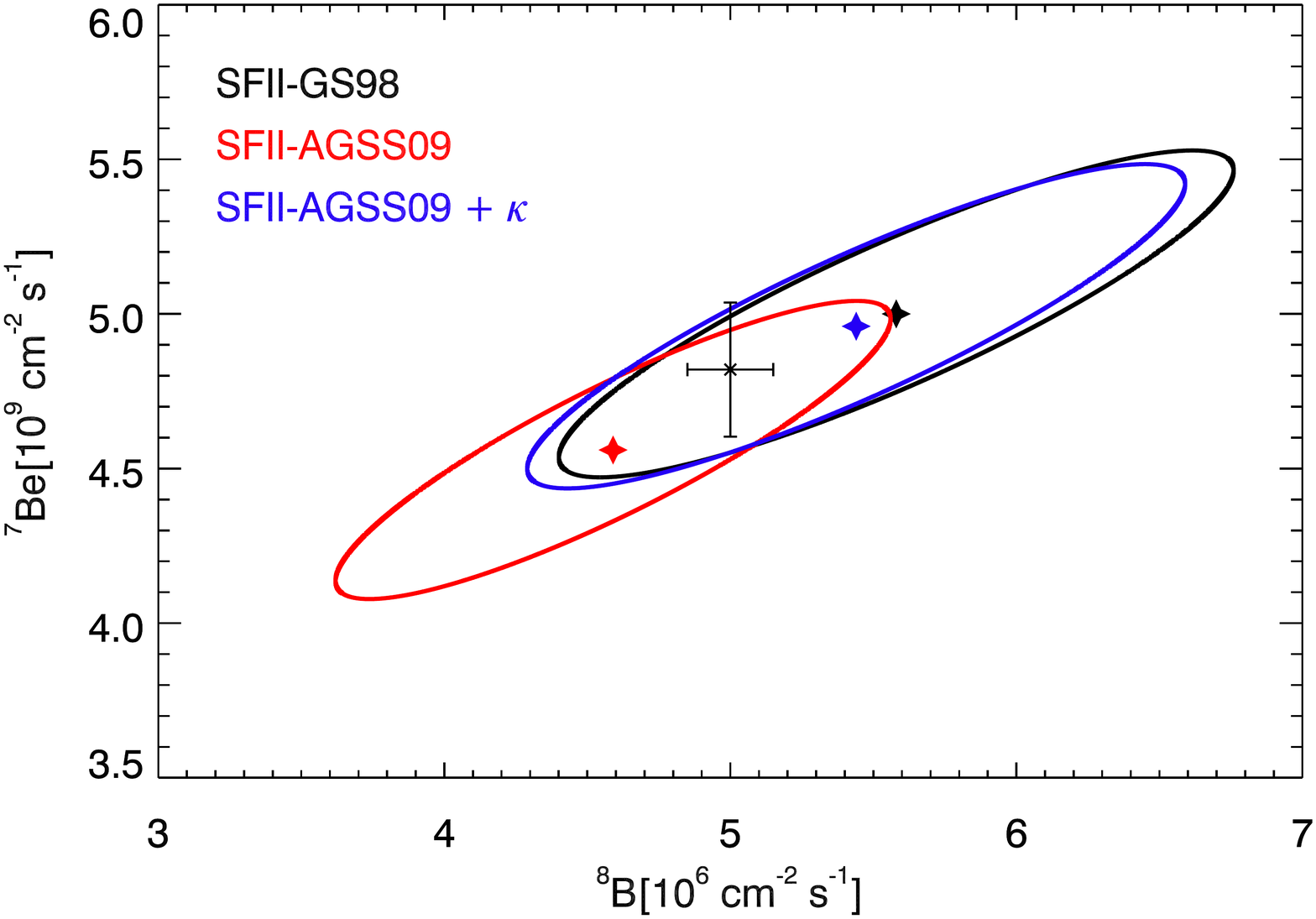}
\qquad\qquad
\includegraphics[height=40mm]{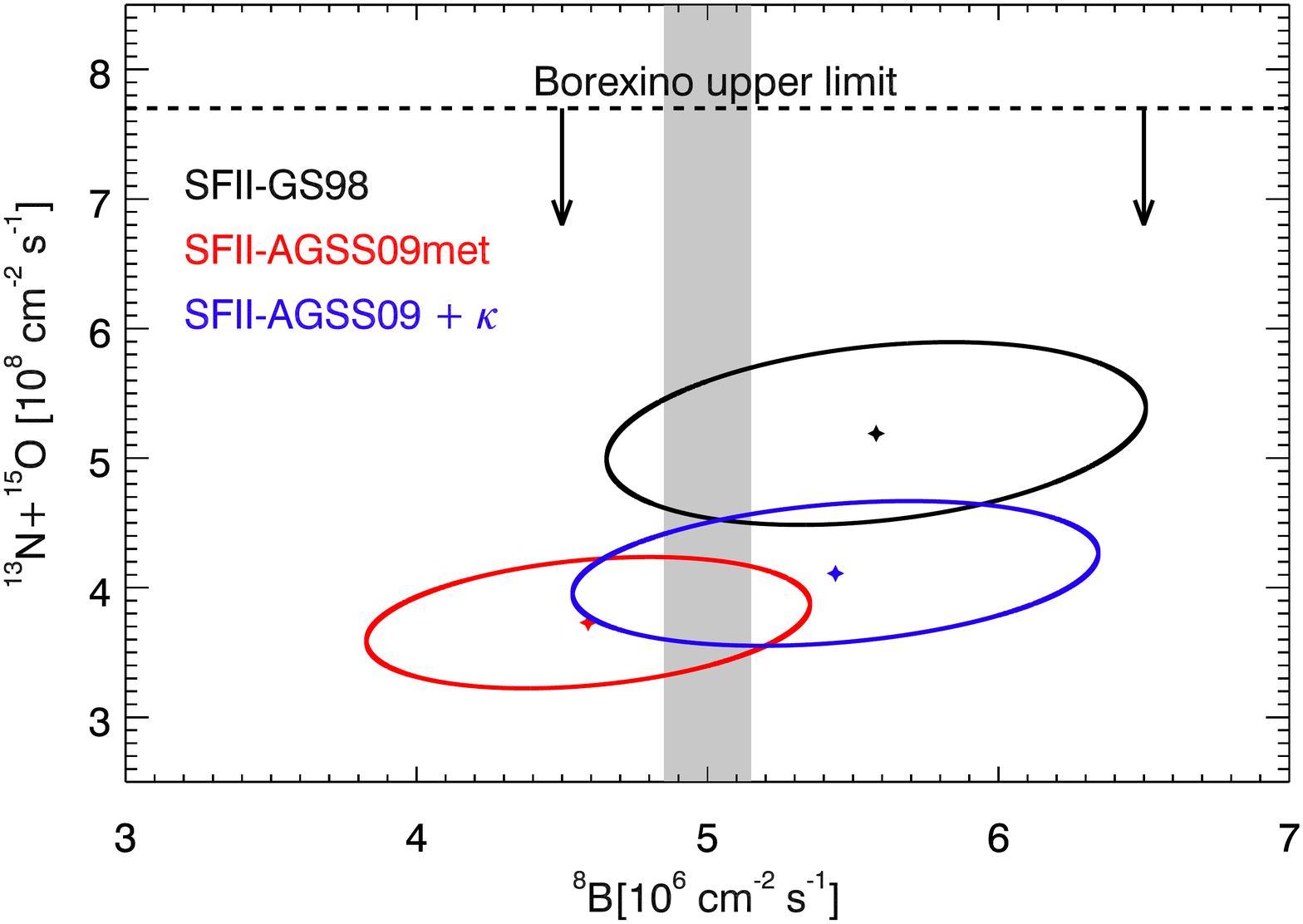}
\caption{
Theoretical predictions and experimental results for solar $\nu$ fluxes. The $^8 {B}$ flux is reported 
on the x-axis; on the vertical axis, instead, one can see the fluxes of $^7{\rm Be}$ (left graph) and of 
$^{13} {\rm N}$ + $^{15}{\rm O}$, from CNO cycle,  (right graph). 
The $1 \sigma$ experimental results correspond to black bars in the left graph and to shaded gray vertical band in the right one. 
The $1 \sigma$ theoretical allowed regions in high Z, low Z and low Z with modified opacity versions of SSM correspond, respectively, to black, red and blue ellipses.Taken from~\cite{Serenelli-graphs}.}
\label{fig-flussi-e-ssm}
\end{figure}

The fig.~\ref{fig-flussi-e-ssm} (left panel) shows the present situation: the experimental results are compatible with the predictions both of 
low Z and of high Z models. It will not be possible to definitely solve this ambiguity only by looking at these two variables and the real breakthrough 
would come by the measurent in future experiments of the CNO neutrino fluxes, as shown in the right panel of fig.~\ref{fig-flussi-e-ssm}. 
However this graph indicates also that a parallel improvement in the determination of one of the two fluxes measurable at JUNO 
(for instance the $^8 {B}$ one) would bring a complementary piece of information, essential to disentangle, for instance, the possible 
ambiguity between high Z models and models predicting low Z metallicity with modified opacity.
 
JUNO should be able to observe $^8{B}$ neutrinos (even in at ``relatively low'' energies),
but the separation of the signal from the noise will be  quite a difficult task, due to the possible presence of  internal 
 and, above all, external background, which must be kept under control. The main problem is linked to cosmogenic background of 
 long lived spallation radioisotopes. The most dangerous for the $\nu_e$ elastic channel are $^8{\rm Li}$, $^{16}{\rm N}$ and, above all,
 $^{11}{\rm Be}$, whose spectrum covers almost all the erergy region of interest.    
 In addition to the usual $\nu_e$ elastic scattering, another, even more interesting, channel to study $^8 {B}$ neutrinos is the 
 charged current interaction, mainly with $^{13} {\rm C}$ ($\nu_e \, + \, ^{13}{\rm C} \to e^- + ^{13}{\rm N}$), with an energy threshold of 
 2.2 MeV.

The hope is that of taking advantage from the high statistics and the excellent energy resolution of JUNO to perform a detailed 
study of the spectrum in the transition region (around 3-4 MeV) between the low energy vacuum oscillation and the higher energies 
where the MSW mechanism dominates (see the first paper of~\cite{metallicity-e-SSM} and the references therein).  
The goal is that of testing the stability of the LMA oscillation solution. The difficulty to observe 
in this region the predicted upturn in the oscillation probability has stimulated a wide discussion in literature and suggested also 
the possibility of considering additional contributions to the ``standard'' oscillation pattern, due to Non Standard Neutrino Interactions.
The models predicting this kind of corrections are of theoretical interest, also because in some cases they can be connected with more 
general extensions of the electroweak Standard Model, including, for instance, some light scalar dark matter candidates. 
JUNO could probably give contributions to the clarification of this topic competitive with the results available from solar neutrino 
experiments (up to now mainly SNO and KamLAND). 

\subsection{Geoneutrinos}
Another topic of great  interdisciplinar interest, not only for physics, but also for geology, is the study of geoneutrino flux~\cite{Bellini:2013wsa}, 
very useful to evaluate the radiogenic contribution to the total heat power of the Earth.
The estimate of our planet heat power ($46 \pm 3 \, {\rm TW}$) is quite accurate, but there is still a great uncertainty on the percentage of 
this quantity due to natural radioactivity.
Measuring the $\bar{\nu}_e$ emitted in the $^{238}{\rm U}$ and $^{232}{\rm Th}$ radioactive decay chains and testing the ${\rm Th/U}$ rate
helps in understanding the abundance of radioactive elements and, therefore, the Earth composition. 
This information should give important insights into the structure of the mantle and the nature of the mantle convection 
(that are still partially unknown) and they could be used to  discriminate between different models on the Earth's 
composition~\cite{geoneutrinos-and-models}, ranging from the so called ``geodynamical models'', that predict a dominant  radiogenic contribution, around 
($33 \pm 3$) TW, to the Earth's heat power (with a value of the Urey ratio for the mantle in the interval between 0.6 and 0.8)\footnote{The Urey ratio for the mantle is defined as the ratio between the radiogenic heat contribution  produced in the mantle and the total heat contribution from the mantle itself.}, to the ``radiochemical models'', in which the 
heat power due to  radioactivity is significantly lower (about $(11 \pm 2)$ TW), with a value of the Urey ratio that can be around 0.1.  

The very interesting JUNO potentiality for the geoneutrino measurements are due to its depth and radiopurity and, above all, to its 
very high size. The main problem is, of course, the presence of a very significant background, 
due to reactor antineutrinos, as shown in fig.7.
Nevertheless, one expects an important geoneutrino measurement at JUNO, which already in the first year of data taking should detect
among 300 and 500 events, that is more than the number of geoneutrinos detected by that time by all the other available experiments 
 (KamLAND, Borexino and SNO+). 
\begin{figure}
\begin{centering}
\includegraphics[scale=0.25]{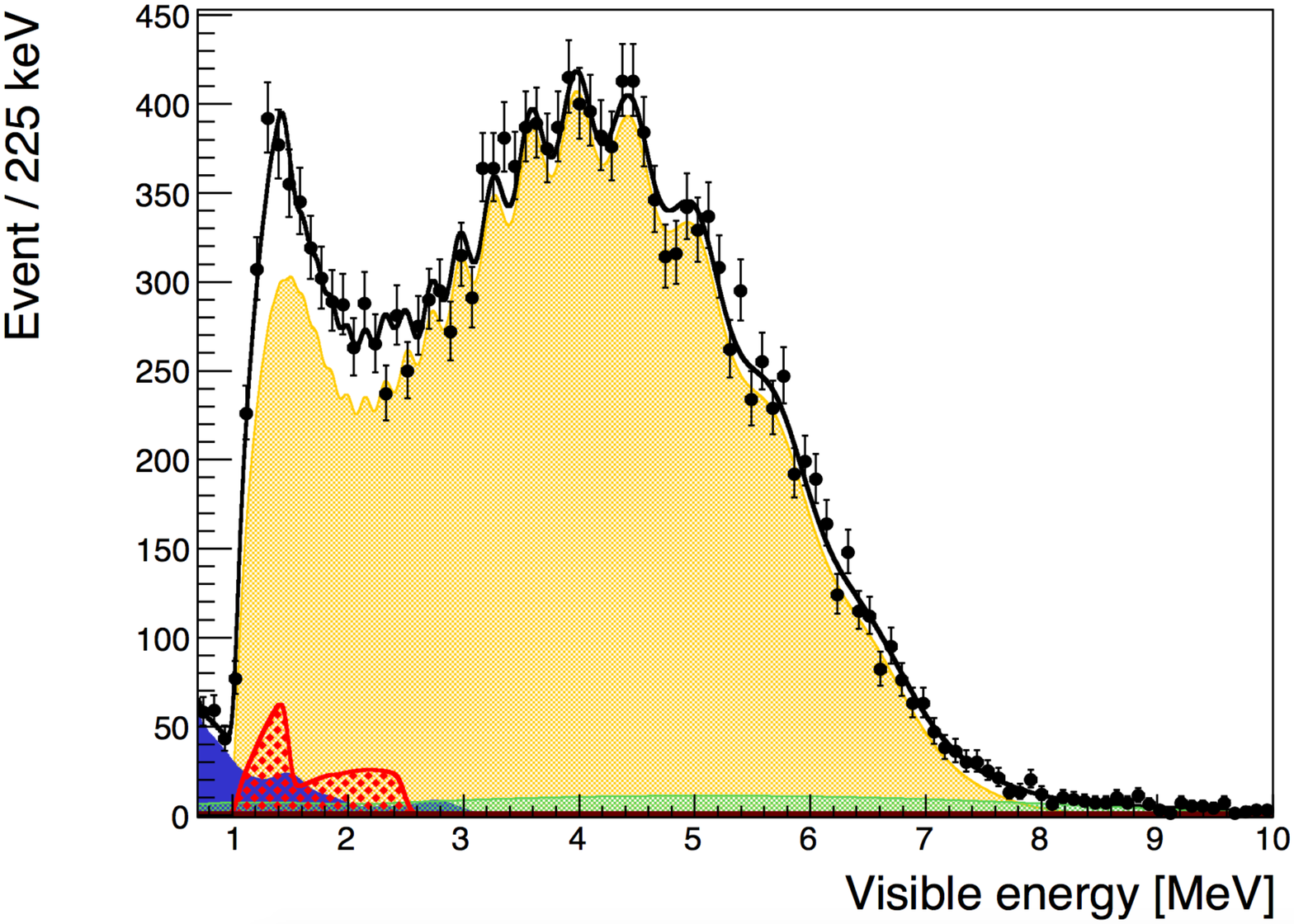}
 \caption{Monte Carlo estimate of the prompt IBD candidates
 for 1 year of measurements, with fixed chondritic Th/U  mass ratio. The expected geo-$\nu$ signal is represented in red and other components are: the reactor antineutrinos (orange), main background component, $^9 {\rm Li}-^8 {\rm He}$ (green), accidental (blue). From~\cite{JUNO-Yellow-Book}.                                                                                                                                                                                                                                                                                                                                                                                                                                                                                                                                                                                                                                                                                                                                                                                                                                                                                                                                                                                                                   
 }
\end{centering}
\label{spettro-geoneutrini}
\end{figure}

\subsection{Other studies at JUNO}
For other interesting measurements will be performed at JUNO, like the measurement of atmospheric neutrinos, the search for sterile neutrinos, 
nucleon decays and dark matter and more exotic searches, we refer the interested reader to~\cite{JUNO-Yellow-Book}. 
%


\begin{thebibliography}{99}
\bibitem{JUNO-Yellow-Book}  
F.~An {\it et al.} [JUNO Collaboration],
  J.\ Phys.\ G {\bf 43} (2016) no.3,  030401.
%
\bibitem{doublebeta}
See, for instance: W.~Rodejohann,
  Int.\ J.\ Mod.\ Phys.\ E {\bf 20} (2011) 1833;
 S.~Dell'Oro, S.~Marcocci, M.~Viel and F.~Vissani,
 Adv.\ High Energy Phys.\  {\bf 2016} (2016) 2162659;
 S.~F.~Ge, W.~Rodejohann and K.~Zuber,
  arXiv:1707.07904 [hep-ph].
\bibitem{double-beta-sterile}
 I.~Girardi, A.~Meroni and S.~T.~Petcov,
 JHEP {\bf 1311} (2013) 146;
S.~Gariazzo, C.~Giunti, M.~Laveder, Y.~F.~Li and E.~M.~Zavanin,
J.\ Phys.\ G {\bf 43} (2016) 033001; 
 C.~Giunti, E.~Zavanin,
 JHEP {\bf 1507} (2015) 171 
 and        
 J.\ Phys.\ Conf.\ Ser.\  {\bf 718} (2016) n.6,  062074. 
 %
\bibitem {LBL-hierarchy}
P.~Adamson {\it et al.} [NOvA Collaboration],
 Phys.\ Rev.\ Lett.\  {\bf 116} (2016) no.15,  151806.
\bibitem{cosmology-hierarchy}
S.~Vagnozzi, {it et al.},
arXiv:1701.08172 [astro-ph.CO].
\bibitem{discussion-hierarchy}
See:  F.~Simpson, R.~Jimenez, C.~Pena-Garay and L.~Verde,
JCAP {\bf 1706} (2017) no.06,  029
and the criticism to it in:
T.~Schwetz {\it et al.}, 
 arXiv:1703.04585 [astro-ph.CO].
\bibitem{Petcov-et-al}
 S.~Choubey, S.~T.~Petcov and M.~Piai,
  Phys.\ Rev.\ D {\bf 68} (2003) 113006;
 and Phys.Lett.B {\bf 533} (2002) 94. 
%
\bibitem{JUNO-technical}
  Z.~Djurcic {\it et al.} [JUNO Collaboration],
  arXiv:1508.07166 [physics.ins-det].
\bibitem{paper-medium-baseline}
  L.~Zhan, Y.~Wang, J.~Cao and L.~Wen,
  Phys.\ Rev.\ D {\bf 78} (2008) 111103\, .
\bibitem{talk-Gioacchino}
Gioacchino Ranucci and JUNO Collaboration,
J.\ Phys.\ Conf.\ Ser.\  {\bf 888} (2017) no.1,  012022.
\bibitem{talk-Yfang}
Y.~Wang, 
``Large area MCP-PMT and its application at the JUNO experiment'', invited contribution to the {\it XVII International Workshop on Neutrino Telescopes}, will appear on PoS (Neutel 017) {\bf 056}.   
\bibitem{theory-chi2}
See~\cite{JUNO-Yellow-Book} and, for instance:
F.~Capozzi, E.~Lisi and A.~Marrone,
  Phys.\ Rev.\ D {\bf 89} (2014) no.1,  013001;
M.~Blennow, P.~Coloma, P.~Huber and T.~Schwetz, 
JHEP {\bf 1403}  (2014) 028;  
S.~F.~Ge, {\it et al.},
JHEP {\bf 1305} (2013) 131;
S.~F.~Ge, K.~Hagiwara, N.~Okamura and Y.~Takaesu,
  JHEP {\bf 1305} (2013) 131;
E.~Ciuffoli, J.~Evslin and X.~Zhang,
  JHEP {\bf 1401} (2014) 095;
L.~Stanco, G.~Salamanna, F.~Sawy and C.~Sirignano,
 arXiv:1707.07651 [hep-ph].
  %
  \bibitem{Petcov-Bilenky}
  S.~M.~Bilenky, F. ~Capozzi and S.~T.~Petcov,
 Phys.\  Lett.\  B  {\bf 772} (2017) 179. 
%
\bibitem{Capozzi-Lisi-plus-Esteban-et-al}
F.~Capozzi, E.~Di Valentino, E.~Lisi, A.~Marrone, A.~Melchiorri and A.~Palazzo,
Phys.\ Rev.\ D {\bf 95} (2017) no.9,  096014; 
F. Capozzi, {\it et al}, 
J.\ Phys.\ Conf.\ Ser.\ {\bf 888} (2017) no.1, 012037; \, 
I.~Esteban, M.~C.~Gonzalez-Garcia, M.~Maltoni, I.~Martinez-Soler and T.~Schwetz,
 JHEP {\bf 1701} (2017) 087. 
\bibitem{metallicity-e-SSM}
V.~Antonelli, L.~Miramonti; C.~Pena Garay and A.~Serenelli, 
Adv.\ High Energy Phys.\  {\bf 2013} (2013) 351926;
W.~C.~Haxton, R.~G.~Hamish Robertson and A.~M.~Serenelli,
  Ann.\ Rev.\ Astron.\ Astrophys.\  {\bf 51} (2013) 21;
  N.~Vinyoles {\it et al.},
  Astrophys.\ J.\  {\bf 835} (2017) no.2,  202.
\bibitem{Serenelli-graphs}
Serenelli, A. M.,  talk given at ``A special Borexino event - Borexino Mini-Workshop - 2014'', Gran Sasso, Sep. 2014; avilable online 
from the site http://borex.lngs.infn.it/ .
\bibitem{Bellini:2013wsa}
 See, for instance: G.~Bellini, A.~Ianni, L.~Ludhova, F.~Mantovani and W.~F.~McDonough,
  Prog.\ Part.\ Nucl.\ Phys.\  {\bf 73} (2013) 1\, . 
\bibitem{geoneutrinos-and-models}
  O.~Sramek, W.~F.~McDonough, E.~S.~Kite, V.~Lekic, S.~Dye and S.~Zhong,
  Earth Planet.\ Sci.\ Lett.\  {\bf 361} (2013) 356\, .
%
\end{thebibliography}
\end{document}